\documentclass[aps,pre,reprint,10pt,superscriptaddress,showpacs]{revtex4-1}
\usepackage{amsfonts} 
\usepackage{amsmath}
\usepackage{amssymb}
\usepackage{graphicx}
\usepackage{dcolumn}
\usepackage{bm}
\usepackage[]{natbib}
\usepackage{subfigure}
\usepackage{color}
\makeatletter
\newcommand*\xbar[1]{%
  \hbox{%
    \vbox{%
      \hrule height 0.5pt 
      \kern0.5ex
      \hbox{%
        \kern-0.1em
        \ensuremath{#1}%
        \kern-0.1em
      }%
    }%
  }%
} 
\begin{document}
\title{Nonlinear evolution of internal gravity waves in the Earth's ionosphere: Analytical and numerical approach}
\author{T. D. Kaladze}
\email{tamaz_kaladze@yahoo.com}
\affiliation{I. Vekua Institute of Applied Mathematics, Tbilisi State University, Georgia}
\affiliation{E. Andronikashvili Institute of Physics, Tbilisi State University, Georgia}
\author{A. P. Misra}
\email{apmisra@visva-bharati.ac.in; apmisra@gmail.com}
\affiliation{Department of Mathematics, Siksha Bhavana, Visva-Bharati University, Santiniketan-731 235,  India}
\author{Animesh Roy}
\email{aroyiitd@gmail.com}
\affiliation{Department of Mathematics, Siksha Bhavana, Visva-Bharati University, Santiniketan-731 235,  India}
\author{Debjani Chatterjee}
\email{chatterjee.debjani10@gmail.com}
\affiliation{Department of Applied Mathematics, University of Calcutta,  Kolkata-700 009, India}
\begin{abstract}
 The nonlinear  propagation of internal gravity  waves in the weakly ionized, incompressible Earth's ionosphere   is studied using the fluid theory approach. Previous theory in the literature is advanced by the effects of the terrestrial inhomogeneous  magnetic field embedded in weakly ionized ionospheric layers ranging in altitude from about $50$ to $500$ km. It is shown that the ionospheric conducting fluids can support the formation of solitary dipolar vortices (or modons).  Both analytical and numerical solutions of the latter are obtained and analyzed.   It is found that in absence of the Pedersen conductivity, different vortex structures  with different space localization can be formed   which can move with the supersonic velocity without any energy loss. However, its presence can cause the amplitude of the solitary vortices to decay with time and    the  vortex structure can completely disappear  owing to the energy loss.  Such energy loss can be delayed, i.e.,  the vortex structure can prevail for relatively a longer time if the nonlinear effects associated with either the stream function or the density variation  become significantly higher than the dissipation due to the Pedersen conductivity. The main characteristic dynamic parameters are also defined which are in good correlation with the existing experimental data.
 \end{abstract}
\maketitle

\section{Introduction}\label{sec-intro}
The internal gravity   waves (GWs) are typically the low-frequency branch of  acoustic gravity waves. Such waves   are stimulated by the buoyancy forces arising from the vertical displacement of atmospheric particles in the gravitational field and are principally connected with the inhomogeneous distribution of  medium density along the vertical direction (density stratification). Typical frequency of these internal GWs varies in  the interval  $10^{-4}$ s$^{-1}<\omega<1.7\times\times10^{-2}$ s$^{-1}$  and the typical wavelength is $\sim10$ km. It should be noted that the internal GWs can propagate only in a stably stratified medium (in which the density of the medium  increases vertically down)  when the square of the Brunt-V{\"a}is{\"a}l{\"a} frequency is positive, i.e., $\omega_g^2>0$. The transfer of wave energy to charged particles in the regions starting from  the Earth's lithosphere to the atmosphere and ionosphere is a fundamental problem of geophysics and applied research. The internal GWs play an important role in such s process. For example, internal GWs, propagating in the upward direction from the Earth's surface to the upper atmosphere and ionosphere, are able to carry a large amount of energy and momentum. These waves are crucial in the atmospheric convection, generation of atmospheric turbulence and may affect global circulation.  Theoretical and experimental studies over the last many years  have shown that the source of acoustic GWs    in the ionosphere can be earthquakes, volcanic eruptions, tornadoes, thunderstorms, solar eclipses, terminators, jet flows, polar and equatorial electrojets, meteors, strong explosions, and powerful rocket launches (see e.g., \citep{kato1980,rottger1981,fovell1992,igarashi1994,grigorev1999,kanamori2004}). 
\par 
The Earth's and solar atmosphere provide a favorable environment for the generation and propagation of internal GWs. The internal GWs and related nonlinear structures are widely observed in the upper and lower layers of the atmosphere as well as in the lower ionosphere. In this context, a comprehensive review of GW dynamics and its effects in the Earth's middle atmosphere has been given by Fritts and Alexander \citep{fritts2003}.  Furthermore,   the important roles of GWs in the  vertical coupling of lower and upper atmospheres have been discussed in  recent reviews \citep{yigit2015,yigit2016}.     The most obvious gravity wave sources and experimental observations are widely discussed in this review (see for relevant details, e.g.,  \citep{swenson1995,picard1998,mitchell1998}). The atmospheric parameters  and the onset of instability  depend on season and altitude. Observations of the mesopause region ($80-105$ km) show that the probabilities of convective and the dynamic instability range  from $1$ to $2$ and from $3$ to $10$ per cent  respectively \citep{sherman2006,zhao2003}.  Klausner \textit{et al.} \citep{klausner2009}  presented an experimental observation of the seasonal variation of gravity waves and the oscillations of traveling ionospheric disturbances  in the ionospheric $F2$ layer during solar activity observed in different levels of geomagnetic activities. Observations of gravity wave characteristics in the Earth's atmosphere were reviewed by Lu \textit{et al.} \citep{lu2009}.  The  observational results of acoustic GWs at ionospheric heights during convective storms were presented by {\v S}indel{\'a}{\~{r}ov{\'a} \textit{et al.} \citep{sindelarova2009}. Incoherent scatter radar power profile observations connected with acoustic GW propagation in the ionospheric $F$-region were discussed by Livneh et al. \citep{livneh2007}. High-resolution radiosonde data was processed by Li \textit{et al.} \citep{li2009}  to obtain the gravity wave variability in the lower stratosphere over the South Pole.   Parameters of internal GWs in the mesosphere/lower thermosphere regions from meteor radar wind data were revealed and described by Oleynikov \textit{et al.} \citep{oleynikov2005} where previous experimental observations were  are also listed.  Recently, Izvekova \textit{et al.} \citep{izvekova2015} had shown the possibility of acoustic GW instability in non-adiabatic terrestrial atmosphere. 
\par 
Recently,   Chatterjee and Misra \citep{chatterjee2021} advanced the nonlinear theory of acoustic GWs in the neutral atmosphere with the effects of the Coriolis force. Roy \textit{et al.} \citep{roy2019} presented the nonlinear dynamical behaviors of acoustic GWs in the neutral atmosphere and possibility of tendency to chaotic states is shown. Kaladze \textit{et al.} \citep{kaladze2018,kaladze2019} studied the properties and peculiarities of linear three-dimensional propagation of electromagnetic internal GWs in an ideally conducting incompressible medium embedded in a uniform magnetic field. It is shown that the ordinary internal gravity waves can couple with Alfv{'e}n waves. The existence of nonlinear acoustic GWs in the form of two-dimensional small-scale vortices (i.e., roll-type structures) in the neutral atmosphere has been reported by Stenflo and Shukla \citep{stenflo2009}. Onishchenko \textit{et al.} \citep{Onishchenko2013} have shown that the presence of finite temperature gradient can substantially modify the condition for the existence of vortex-type structures. It was noted that the vortices of internal GWs with the spatial scale $L$ smaller than the reduced height $H$ can be originated only in a convective unstable atmosphere. The amplitude of internal GWs propagating upward in various regions of the Earth's atmosphere and ionosphere can grow with height and become unstable due to the parametric instability associated with the  resonant wave-wave interactions. Such parametric generation of large scale zonal structures by high-frequency finite-amplitude internal GWs has been investigated recently by Horton \textit{et al.} \citep{horton2008} and Onishchenko  \textit{et al. } \citep{onishchenko2012}. The effects of the wind shear on the roll structures of nonlinear internal GWs   in the Earth's dynamically unstable atmosphere with the finite vertical temperature gradients were investigated by Onishchenko et al. \citep{Onishchenko2014}.
\par 
Theoretical investigation of the influence of charged particles and the inhomogeneous terrestrial magnetic field on the possibility of the formation of solitary vortical structures in  different layers of the Earth's weakly ionized ionosphere was started by Kaladze \textit{et al.} \citep{kaladze1997}. The propagation of electromagnetic inertio-gravity waves in the earth's partially ionized ionosphere was undertaken by Kaladze \textit{et al.} \citep{kaladze2007,kaladze2011a}.  The propagation characteristics of acoustic GWs  in  weakly ionized ionospheric $D$-, $E$-, and $F$-layers were investigated by Kaladze \textit{et al.} \citep{kaladze2008a}. It was shown  that the incorporation of the Earth's rotation provides a substantial change in the propagation dynamics of low-frequency internal GWs, namely, the existence of the new cut-off frequency at $2\Omega_0$ (where $\Omega_0$ is the value of the angular velocity of the Earth's rotation) is noted. In addition, the authors noted that in the $E$  and $F$ layers of the ionosphere these waves decay with the damping rate of the same order of magnitude as in the linear case. Furthermore, the propagation of  electromagnetic internal GWs coupled to  Alfv{\'e}n waves  in the Earth's ionospheric $E$-layer was investigated by Kaladze \textit{et. al.} \citep{kaladze2011b}. Also, the nonlinear generation of low-frequency and large-scale zonal flow and   relatively small-scale electromagnetic coupled internal gravity and Alfv{\'e}n waves in the Earth's stratified weakly ionized ionospheric $E$-layer was investigated by Kaladze \textit{et al.} \citep{kaladze2012}. The influence of the DC electric field on the heating and instability of internal GWs leading to the formation of solitary vortex structures was studied by Chmyrev and Sorokin \citep{chmyrev2010}. The nonlinear acoustic GWs, generated by  seismic activity, propagating in the stable stratified ionospheric $E$-layer   were shown \citep{kaladze2008b} to cause intensification of atomic oxygen green line emission when their amplitudes become sufficiently large.
On the other hand, the importance of GWs in the global circulation of the middle and upper atmospheres has been recognized through various  observational and theoretical studies.  Using a general circulation model (GCM),  the global distribution and the characteristics of GWs in these   environments have been examined by a number of authors   \citep{miyoshi2008,sato1999}   to investigate the evolution of GWs and their momentum fluxes. 
 In this context, different physical effects have been included in GCM to find estimates that conform with observational results. For example, the GW drag   was included in the GCM and numerical prediction model to simulate a realistic stratospheric weak wind layer   through various parameterizations \citep{iwaski1989}. Furthermore, using a nonlinear GW scheme, the variability of GW effects on the zonal mean circulation has been studied with the middle and upper atmospheric models  \citep{lilienthal2020}. Miyoshi \textit{et al.}  \citep{miyoshi2008} studied the characteristics of GWs in the mesosphere and
thermosphere using a GCM that includes the global distributions of electrons and ions.  However, they commented that the model may not be appropriate to predict the ionospheric fluctuations  induced by the neutral atmospheric variation.   
\par 
In this paper, our aim is to advance the work of Kaladze \textit{et al.} \citep{kaladze2008a} in investigating the existence and the nonlinear evolution of different kinds of solitary vortices that can be formed in the propagation of  internal GWs in different layers of the Earth's ionosphere. Using the fluid theory approach, a set of nonlinear coupled equations for the stream function and the density perturbation of atmospheric fluids is derived which governs  the   evolution of nonlinear solitary vortices. Both analytical and numerical studies are performed to obtain and analyze different kinds of stationary propagating  dipolar vortical solutions as well as their  dynamical evolution with the effects of Pedersen conductivity and the system nonlinearities.  The manuscript is organized as follows. In Sec. \ref{sec-model}, the physical and mathematical modelings of internal GWs are described. While Sec.  \ref{sec-analyt}  is devoted to study analytically the existence of different kinds of stationary dipolar solitary vortices, the dynamical evolution of such dipolar vortices is studied numerically in Sec. \ref{sec-numer}. Finally, Sec. \ref{sec-conclu}  is left for discussion and concluding our results. 
\section{Physical and mathematical modeling} \label{sec-model}
 We consider the nonlinear propagation of internal gravity waves in the Earth's  ionospheric layers ($100-500$ km heights)  stratified by the gravitational field. We assume that the {ionospheric layers are weakly ionized and they consist  of electrons, positive ions, and a bulk of massive neutral particles  (i.e., $n/N\ll1$, where $n$ and $N$, respectively, denote the equilibrium number densities of   electrons or ions and neutrals). So, the presence of  charged particles make the ionosphere to be electrically conducting.    Due to strong collisions between the ionized and neutral particles, the dynamics of such electrically conducting ionosphere significantly differ  from that of neutral atmosphere.  Further we assume that  the ionospheric plasma is   immersed in the geomagnetic field $\mathbf{B}_0$ and is under the influence of the Coriolis force  due to the Earth's rotation with the angular velocity $\mathbf{\Omega}_0=(0,0,\Omega_0)$. It follows that the interaction of the induced ionospheric current ($\mathbf{j}$) with the geomagnetic field ($\mathbf{B}$), i.e., the  influence of the Amp{\'e}re force ($\mathbf{j}\times\mathbf{B}$-force)   should also be taken into consideration.    
 \par 
It is to be noted that  for ionospheric disturbances the effective magnetic Reynolds'   number is relatively small in the ionospheric $D$, $E$, and $F$ regions, i.e.,   $R_\text{eff}\approx \mu_0\sigma_\text{eff}VL\sim VL/ \eta_m \ll1$, where $\mu_0=4\pi\times10^{-7}$ H/m is the permeability of free space, $\eta_m$ is the magnetic diffusivity or plasma resistivity,  $\sigma_\text{eff}\sim\sigma_P+\sigma_H^2/\sigma_P$ is the effective conductivity ($\sim1/\eta\mu_0$)   of the ionosphere  with $\sigma_H$ denoting the Hall conductivity; and $V$ and $L$ are, respectively, the characteristic dimensions of velocity and space.  So, for the lower ionosphere, one can neglect the induced magnetic field  $\mathbf{b}=R_\text{eff}\mathbf{B}$ and the  vortex part of the self-generated electric field   $E_v\sim R_\text{eff} V B$ which originates due to the fluctuating magnetic field $\mathbf{b}$. Thus, for wave-like perturbations in the ionosphere we can consider the magnetic field as specified by the external geomagnetic  field only, i.e.,  $\mathbf{B}=\mathbf{b}+\mathbf{B}_0\approx \mathbf{B}_0$ and $E_v\approx0$ such that $\nabla\cdot\mathbf{B}_0=0$ and $\nabla\times\mathbf{B}_0=\mathbf{0}$. In this non-inductive approximation, it is sufficient to consider only the current $\mathbf{j}$ originating in the medium \citep{kaladze2008a} and  the action of the geomagnetic field $\mathbf{B}_0$ on the induction current $\mathbf{j}$ in ionospheric plasmas makes it necessary to take into consideration the Amp{\'e}re force  in the equations for ionospheric dynamics. The most important impact of this force is the appearance of the inductive (magnetic) damping (due to the Pedersen currents) in the Earth's ionosphere, which is no less significant than viscous breaking, especially in the $F$ region.
\par 
In what follows, the dynamics of the electrically conducting ionospheric plasmas can be described by the following momentum equation.
 \begin{equation}
 \frac{\partial \mathbf{u}}{\partial t}+\mathbf{u}\cdot\nabla\mathbf{u}=-\frac{\nabla p}{\rho}+\frac{\mathbf{j}\times\mathbf{B}_0 }{\rho}+\mathbf{g}, \label{eq-moment}
 \end{equation}
where $\mathbf{j}$ is the  current density, $\mathbf{u}$ is the bulk (neutral) velocity; $p$ and $\rho$ are the pressure and mass density of the medium, and $\mathbf{g}=(0,0,-g)$ is the  acceleration due to gravity. Equation \eqref{eq-moment} is  supplemented by the continuity  equation and the equation of state, given by, 
 \begin{equation}
 \frac{\partial \rho}{\partial t}+\nabla\cdot(\rho\mathbf{u})=0, \label{eq-cont}
\end{equation}
\begin{equation}
\left(\frac{\partial }{\partial t}+  \mathbf{u} \cdot \nabla \right)\left(p-  c_s^2 \rho\right) =0, \label{eq-press}
\end{equation}
where $c_s$ is the sound speed. The background pressure $p_0$ and the mass density $\rho_0$ are stratified by the gravitational field. They  can be assumed to vary as $\rho_0(z)=\rho_0(0)\exp(-z/H)$ and $p_0(z)=p_0(0)\exp(-z/H)$,  where   $H=c_s^2/\gamma g$ is the  reduced scale length of the   atmosphere   with $\gamma$ denoting the adiabatic constant.   
Since the ionosphere can be considered as quasineutral with a high degree of accuracy, we can ignore the inner electrostatic field, i.e., $\mathbf{E}=-\nabla\phi=\mathbf{0}$, where $\phi$ is the electrostatic potential, as well as the vortex  components of the self-generated electromagnetic field. Thus, in the conducting medium, the electric field strength is determined only by the dynamo field $\mathbf{E}_d=\mathbf{u}\times\mathbf{B}_0$  and the generalized Ohm's law reduces to
\begin{equation}
\mathbf{j}=\sigma_\perp\mathbf{E}_{d\perp}+\frac{\sigma_H}{B_0}\mathbf{B}_0\times \mathbf{E}_d, \label{eq-j}
\end{equation}
where $\sigma_\perp\equiv\sigma_p$ and $\sigma_H$, respectively, denote the perpendicular (Pedersen)  and Hall conductivities. The  subscripts $\parallel$ and $\perp$ denote the components parallel and perpendicular to the external magnetic field. 
\par 
Next, we introduce a local Cartesian   coordinate system  with the $x$-axis directed from  west to  east, the $y$-axis from   south to   north, and the $z$-axis along the local vertical direction.
We are primarily interested in the dynamics at high latitudes in the Northern Hemisphere, assuming that the geomagnetic field  is vertical and downward, i.e., $\mathbf{B}_0=-B_0\hat{z}$. We  consider the high-latitudes of the Northern Hemisphere to find evident analytic solutions. Also,  in this region,  the tropospheric flow is known to be much less zonal and the topography barriers are much less significant in the north–south orientation  than those in the Southern Hemisphere \citep{garcia2017}.  Furthermore, for the internal GWs   we consider $\mathbf{B}_0$ to be uniform and ignore the influence of the  Coriolis force \citep{kaladze2008a}. Also, to exclude the high-frequency acoustic mode we make use of the  incompressibility condition, given by,
\begin{equation}
\nabla\cdot\mathbf{u}=0. \label{eq-incompr}
\end{equation}
Equations \eqref{eq-moment} to \eqref{eq-incompr} constitute a full set of equations necessary for the description of   vortex motions of  low-frequency incompressible internal GWs.  For simplicity, we consider the two-dimensional motion of internal GWs in the  $xz$-plane by assuming  $\partial/\partial y\sim0$. The incompressibility  condition  \eqref{eq-incompr} allows us to introduce the stream function $\psi$, given by,
\begin{equation}
u=-\frac{\partial\psi}{\partial z},~~w=\frac{\partial\psi}{\partial x}
\end{equation}
and the  $y$-component of the   vorticity vector $\nabla\times\mathbf{u}$ with $\mathbf{u}=(u,0,w)$   as
 \begin{equation}
 \zeta=\frac{\partial u}{\partial z}-\frac{\partial w}{\partial x}=-\left(\frac{\partial^2}{\partial x^2}+\frac{\partial^2}{\partial z^2}\right)\psi\equiv-\nabla^2\psi. \label{eq-zeta}
\end{equation}                                                                                                           Following  the work of Kaladze \textit{et al.}  \citep{kaladze2008a}, we obtain the following   system of equations which describe the nonlinear evolution of low-frequency internal GWs in the weakly ionized stable stratified Earth's ionosphere. 
\begin{equation}
\begin{split}
 \frac{\partial}{\partial t}\left(\nabla^2 \psi -\frac{\psi}{4H^2} \right) &+J(\psi, \nabla^2 \psi) =-\frac{\partial \chi}{\partial x} \\
 &-\frac{\sigma_p B_0^2}{\rho_0}\left(\frac{\partial^2 \psi}{\partial z^2} -\frac{\psi}{4H^2} \right), \label{eq-psi} 
 \end{split}
\end{equation}
 \begin{equation}
 \frac{\partial \chi}{\partial t} +J(\psi,\chi)=\omega_g^2 \frac{\partial \psi}{\partial x}, \label{eq-chi}
 \end{equation}
where $\omega_g=\sqrt{g/H}$ is the Brunt-V{\"a}is{\"a}l{\"a} frequency for the incompressible fluid,    the nonlinear term $J$ is the Jacobian,  given by,  
\begin{equation}
J(a,b)=\frac{\partial a}{\partial x}\frac{\partial b}{\partial z}-\frac{\partial a}{\partial z}\frac{\partial b}{\partial x},
\end{equation}
 and $\chi=g\rho/{\rho_0(0)}$  with $\rho$ denoting the perturbed mass density.  Also,      $\eta_0=\sigma_pB_0^2/\rho_0$ is the Pedersen parameter  which produces the magnetic damping  of internal GWs.
 From  Eqs.  \eqref{eq-psi} and \eqref{eq-chi} it is straightforward to obtain the following energy conservation equation.  
  \begin{equation}
  \frac{\partial {\cal E}}{\partial t}+\eta_0 \int\int \left[\left(\frac{\partial \psi}{\partial z}\right)^2+\frac{1}{4}\frac{\psi^2}{H^2} \right]dxdz=0, \label{eq-enr1}
  \end{equation}
where $\eta_0>0$ and ${\cal E}$ is the energy of the solitary vortex of internal GWs, given by, 
 \begin{equation}
{\cal E}=\int\int\left[\frac{1}{2}\left(\nabla\psi\right)^2+\frac{1}{8}\frac{\psi^2}{H^2}+\frac{1}{2}\frac{\chi^2}{\omega_g^2}\right]dxdz. \label{eq-ener2}
\end{equation}
Since the second term in Eq. \eqref{eq-enr1}  is positive definite, it follows from the $H$-theorem, 
 stated as,   If a   distribution function of a particle $f(\mathbf{r},\mathbf{v},t)$ satisfies the Boltzmann transport equation,   then the system evolves in such a manner that $dH/dt\leq0$, and if $dH/dt=0$, then the system is in the equilibrium state, where $H$ is the functional, given by,
\begin{equation}
H(t)=\int f(\mathbf{r},\mathbf{v},t) \ln f(\mathbf{r},\mathbf{v},t)d^3r d^3v, 
\end{equation}  
that $\partial {\cal E}/\partial t\leq0$. Thus, the wave energy decays with time by the effects of the Pedersen parameter $\eta_0$ and so a steady-state solution with finite wave energy does not exist.   Now, as per the  estimations provided by Kaladze \textit{et al.} \citep{kaladze2021}, we have in the ionospheric $E$-layer  the density ratio   $n/N\sim10^{-8}-10^{-6}$, the ion-neutral collision $\nu_{in}\lesssim10^3~\mathrm{s}^{-1}$ so that the parameter $\eta_0\equiv\sigma_pB_0^2/\rho_0\sim\nu_{in}(n/N)$  varies in the interval $(10^{-8}-10^{-5})$ s$^{-1}$, while in the $F$-layer we have $n/N\sim10^{-5}-10^{-3}$,  $\nu_{in}\lesssim10~\mathrm{s}^{-1}$ for which $\eta_0$ varies in  $(10^{-6}-10^{-4})$ s$^{-1}$.   Taking into account this estimation, we look for the linear solution of the system of Eqs. \eqref{eq-psi} and \eqref{eq-chi}. To this end,  we assume the perturbations to vary as plane waves $~\exp(ik_xx+ik_zz-i\omega t)$ with frequency $\omega$ and wave number $k$, and $\omega=\omega_0+i\gamma_0$, where $|\gamma_0|\ll\omega_0$. Thus, we obtain  the following dispersion relation for the internal GWs (which is also valid for the ionospheric neutral $D$-layer).  \citep{kaladze2008a}
 \begin{equation}
 \omega_0^2=\frac{k_x^2\omega_g^2}{k_x^2+k_z^2+1/4H^2}, \label{eq-omega0}
 \end{equation}
 and the wave damping rate (valid for $E$ and $F$ layers) as
 \begin{equation}
 \gamma_0=-\frac{1}{2}\eta_0\frac{k_z^2+1/4H^2}{k_x^2+k_z^2+1/4H^2}. \label{eq-gamma}
 \end{equation}
 \section{Evolution of internal gravity solitary vortex: Analytical approach} \label{sec-analyt}
The nonlinear behaviors of the low-frequency internal gravity waves is greatly influenced by the presence of the advective derivative. The corresponding vector nonlinearity (the Jacobian) can thus produce various coherent localized vortex structures for a broad range of background configurations.  The forms of such vortices are strongly dependent on the spatial profile of the unperturbed medium.  In a quiescent neutral atmosphere with the  exponentially varying density  and pressure profiles, the standard (spatially exponentially localized) travelling dipolar vortices (also called the internal gravity modons) have been found in an unstable stratified neutral $(\sigma_p=0)$ atmosphere with the transverse dimensions either much smaller \citep{stenflo1987}, or comparable with the density scale length in a stable stratified neutral atmosphere \citep{stenflo1995}.  It has been shown that such localized vortex structures can evolve in the form of circular vortices in inner $(r<a)$ and outer $(r>a)$ regions with $a$ denoting the radius of the circle \citep{stenflo1987,shukla1998}. Furthermore, by a numerical simulation approach it has been established that the dipoles associated with anti-cyclones in GWs can become  more circular than cyclonic vortex and spread over a wide region of space \citep{snyder2007}.   
\par
From Eq. \eqref{eq-omega0} we conclude that the phase velocity $V_\text{ph}=\omega/k_x$ is limited by the interval
\begin{equation}
 -V_\text{max}\leq V_\text{ph} \leq V_\text{max}, \label{eq-vmax}
\end{equation}
   where in the case of the incompressible atmosphere,   $V_\text{max}\sim2H\omega_g=2\sqrt{gH}=2c_s/\sqrt{\gamma}$.    Thus, when a source moves along the $x$-direction with a velocity larger than $V_\text{max}$, there is no resonance with the internal GWs. This means that such waves will not be generated by the source and hence no energy loss   \citep{kaladze2008a}. In this situation, one can obtain a stationary solution for the localization of a pulse propagating horizontally with a velocity $|V|>V_\text{max}$. An estimation with the reduced atmospheric height,   $H\sim 6$ km  gives $H\omega_g\sim250$ m/s, i.e.,  $V_\text{max}\sim 500$ m/s. It follows that  the  nonlinear solitary vortex structure so formed   becomes supersonic and its amplitude does not decay due to the generation of linear waves in the region $|V|<V_\text{max}$.
 \subsection{Standard stationary propagating internal gravity dipole vortical solution with $\eta_0\sim0$ } \label{sec-vortex-sol1}
We note  that in contrast to the equations in \citep{stenflo1987}   and \citep{stenflo1995}, the nonlinear system of Eqs. \eqref{eq-psi} and \eqref{eq-chi}  are obtained under the incompressibility condition \eqref{eq-incompr}, allowing to exclude the high-frequency acoustic gravity mode. To find the regular at the center and exponentially vanishing at  infinity undamped  travelling wave solution for $\eta_0\sim0$, we consider the formation of internal gravity vortices in a frame moving with the constant velocity  $U$ along the $x$-axis and thus apply  the transformations  $\xi=x-Ut$ and $z=z$.   Furthermore, we consider  the following linear relation between the mass density and the stream function. 
 \begin{equation}
 \chi=-\frac{\omega_g^2}{ U}\psi. \label{eq-chi-sol1}
 \end{equation}
 Substituting the relation \eqref{eq-chi-sol1} into Eq. \eqref{eq-psi}, introducing a circle with radius $r=a$,   dividing the integral domain   in the inner $(r<a)$ and the outer $(r>a)$ regions, and following  references \citep{stenflo1987}  and \citep{stenflo1995}, we obtain the following dipole type vortex solutions in the polar coordinates $\xi=r\sin\phi,~z=r\cos\phi$.
 \begin{equation}
 \begin{split}
 &\nabla^2\psi_i=-Ak^2J_1(kr)\cos\phi,~~(r<a),\\
\mathrm{i.e.,}~ &\psi_i=\left[AJ_1(kr)-\frac{k^2+p^2}{k^2}Ur\right]\cos\phi, ~~(r<a) 
  \end{split}
\end{equation}   
 \begin{equation}
 \begin{split}
 &\nabla^2\psi_o=Bp^2K_1(pr)\cos\phi,~~(r>a),\\
 &~\mathrm{i.e.,}~\psi_o=BK_1(pr)\cos\phi,  ~~(r>a) 
 \end{split}
\end{equation}   
where 
\begin{equation} 
p^2=\frac{1}{4H^2}-\frac{\omega_g^2}{U^2}>0. \label{eq-p2}
\end{equation} 
 Also, $J_n$ and $K_n$, respectively, stand for the Bessel function of first kind of order $n$ and   the McDonald  function  of order $n$.                     
In order that the circle $r=a$ to be a streamline we must have for the constant coefficients  
 \begin{equation}
 A=\frac{p^2 U a}{k^2 J_1(ka)},~~B=-\frac{U a}{K_1(pa)}. \label{eq-A-B}
 \end{equation}
Furthermore, the continuity of $\partial \psi/\partial r$ on the circle $r=a$ requires the following parameter matching condition.
 \begin{equation}
\frac{J_2(ka)}{kaJ_1(ka)} =-\frac{K_2(pa)}{paK_1(pa)},
 \end{equation}
which connects the parameters $k$, $p$, and $a$. Without loss of generality,  any two of these parameters, say $k$ and $p$ can  be taken as independent ones. So, two other parameters $U$ and $a$    still remain undefined in the solution. 
Thus, a stationary propagating nonlinear vortical solution for internal GWs  can be written as
 \begin{equation}
 \psi(r,\phi)=aUF(r)\cos\phi, \label{eq-psi-sol}
 \end{equation}
where 
\begin{equation}
F(r)=\left\lbrace\begin{array}{cc}
\frac{p^2}{k^2}\frac{J_1(kr)}{J_1(ka)}-\frac{(p^2+k^2)r}{ak^2},& (r<a)\\
-\frac{K_1(pr)}{K_1(pa)}, & (r\ge a)
\end{array}\right.
\end{equation}                                                                                              
We note that the   solution \eqref{eq-psi-sol} is antisymmetric with respect to the coordinate $z$. Also, the vorticity of velocity $\nabla^2\psi$ is continuous at $r=a$. The vortex solutions thus obtained are  called  dipole vortices. In addition, according to the condition $p^2>0$ [Eq. \eqref{eq-p2}], the vortical structure can propagate in both the positive and negative directions along the $x$-axis with the supersonic velocity $U$, satisfying the condition $U^2>V_\text{max}^2$. Thus, the velocity of the vortical flow $U$ has the values outside the regime $|V_\text{ph}|<V_\text{max}$ which is the regime  of the phase velocity   of linear periodic waves discussed before. So, there is no resonance with linear waves, implying that the  vortices propagate quicker than linear waves due to the nonlinear amplitude. This means that these waves will not be produced by the source with no energy loss.
\par 
A profile of the vortex solution of Eq. \eqref{eq-psi-sol} is shown in Fig. \ref{fig:analytic1}:  (a) in the inner region $r<a$ where the dipolar vortex is formed and (b) in the outer region $r\geq a$ where no such dipole appears.   The typical size of the vortex $a$ can be estimated by using Eq. \eqref{eq-p2} as
 \begin{equation}
 a\sim p^{-1}=\frac{2H|U|}{\left(U^2-4H^2\omega_g^2\right)^{1/2}}\approx 2H. \label{eq-esti-a}
\end{equation}                                                                                                                     
It is worthwhile to mention that  the linear dispersion relation for internal GWs [Eq. \eqref{eq-omega0}]   can be recovered from Eq. \eqref{eq-p2}  by the formal substitution:
 \begin{equation}
 p^2\rightarrow -(k_x^2+k_z^2), ~U\rightarrow\frac{\omega_0}{k_x}.
\end{equation}   
\begin{figure*}! 
\begin{center}
\includegraphics[scale=0.4]{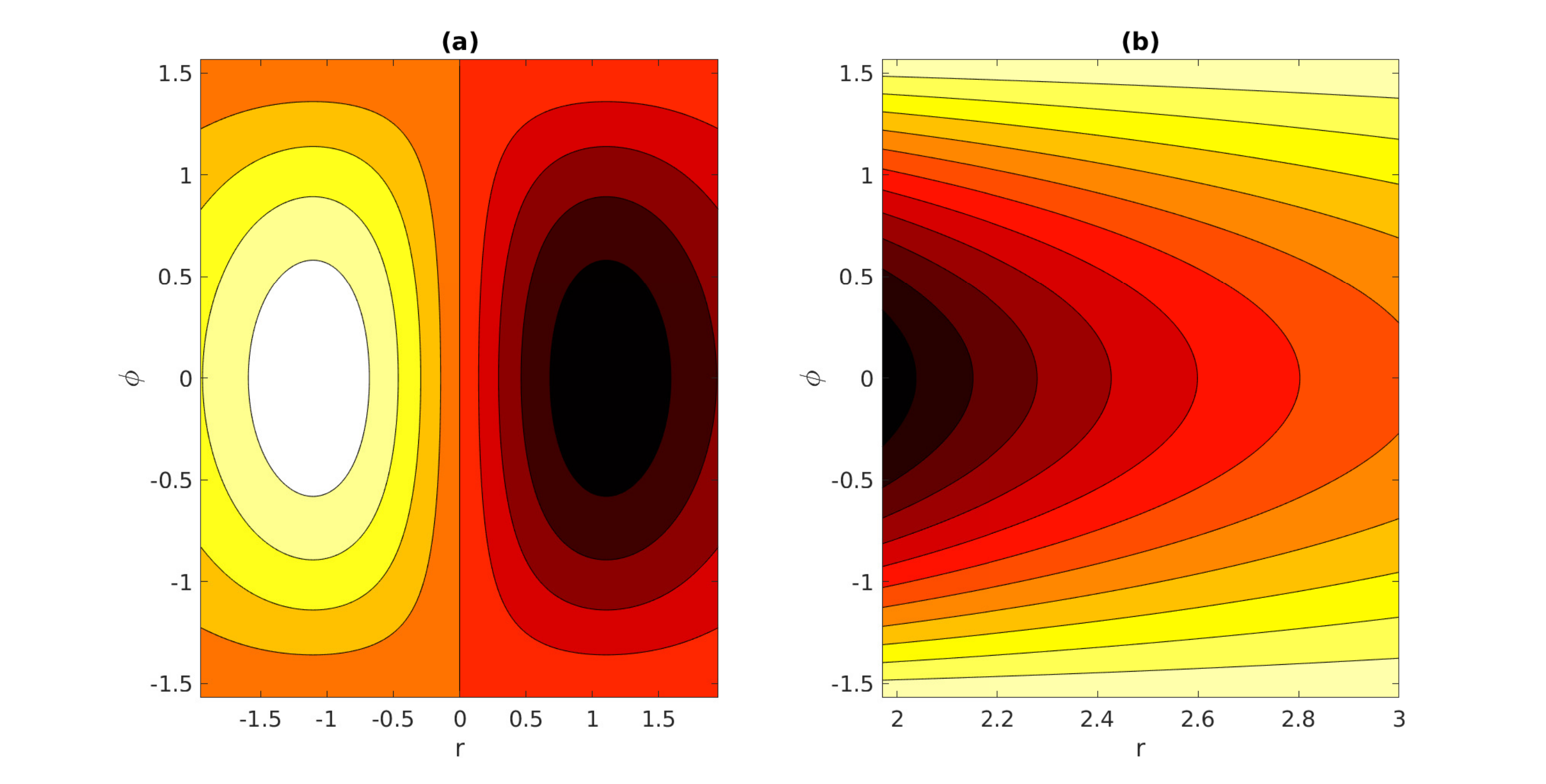}
\caption{Contour plots of  the stationary propagating vortical solution [Eq. \eqref{eq-psi-sol}] are  shown in the $r\phi$ plane for (a) $r<a$ and (b) $r\geq a$. The parameter values are $U/H\omega_g=3,~a/H=1.96,~p\sim a^{-1}$ and $k=2/H$.}
\label{fig:analytic1}
\end{center}
\end{figure*}
\subsection{Other stationary propagating internal gravity dipole vortical solution with $\eta_0\sim0$}   \label{sec-vortex-sol2}
 In this subsection,  we  find another vortex solution of Eqs. \eqref{eq-chi} and \eqref{eq-psi}   by considering the following relation [instead of  Eq. \eqref{eq-chi-sol1}].
 \begin{equation}
 \chi=-\omega_g H\nabla^2\psi. \label{eq-chi-sol2}
\end{equation} 
In this case, Eqs. \eqref{eq-psi} and \eqref{eq-chi} reduce to 
\begin{equation}
\frac{1}{4H^2}\frac{\partial \psi}{\partial t}+\omega_g H\frac{\partial }{\partial x}\nabla^2\psi+\frac{\omega_g}{H}\frac{\partial \psi}{\partial x}=0. \label{eq-psi11}
\end{equation}
Transforming Eq. \eqref{eq-psi11} into the moving frame of reference with coordinates $\xi=x-U t$ and $z=z$, we obtain
\begin{equation}
\frac{\partial}{\partial\xi}\nabla^2\psi+\frac{1}{H^2}\left(1-\frac{1}{4}\frac{U}{H\omega_g}\right)\frac{\partial\psi}{\partial\xi}=0. \label{eq-psi22}
\end{equation}
An integration of Eq. \eqref{eq-psi22} with respect to $\xi$ yields
\begin{equation}
\nabla^2\psi+\frac{1}{H^2}\left(1-\frac{1}{4}\frac{U}{H\omega_g}\right)\psi=F(z), \label{eq-psi33}
\end{equation} 
where $F(z)$ is an arbitrary  function of $z$. Next, looking for a solution of  Eq. \eqref{eq-psi33}, which vanishes at infinity $[F(z)=0]$,  we get 
\begin{equation}
\psi=AJ_1(kr)\cos\phi, \label{eq-psi-sol3}
\end{equation}
where $A$ is a constant and 
\begin{equation}
k^2=\frac{1}{H^2}\left(1-\frac{1}{4}\frac{U}{H\omega_g}\right)=\frac{1}{H^2}\left(1-\frac{1}{2}\frac{U}{V_\text{max}}\right)>0. \label{eq-k2}
\end{equation}
Now, as per Eq.  \eqref{eq-vmax}, the positive velocity $U~(>0)$ should be larger than $V_\text{max}$  and    it varies in the narrow interval
$V_\text{max}<U<2V_\text{max}$. Thus, it follows that the solitary vortex with only the negative velocity can exist having the profile given by Eq. \eqref{eq-psi-sol3}. It is also noted that the dipole vortex solution is regular at the centre   $r=0$  and its amplitude $\propto1/\sqrt{kr}$, i.e., vanishes at infinity. The vortex size can be estimated as 
$a\sim k^{-1}\sim H$. 
\par 
In the case of large $U~(>0)$, Eq. \eqref{eq-psi33} with $F(z)=0$ has the following solution.
\begin{equation}
\psi=BK_1(pr)\cos\phi, \label{eq-psi-sol4}
\end{equation}
where $B$ is a constant and 
\begin{equation}
p^2=\frac{1}{H^2}\left(\frac{1}{4}\frac{U}{H\omega_g}-1\right)>0,
\end{equation}
the positive velocity $U~(>4H\omega_g)=2V_\text{max}$, and the negative velocities are inhibited. We note that the solution \eqref{eq-psi-sol4} has the singularity at the center $r=0$, however, it is exponentially localized in space $[\propto\exp(-pr)]$. Typical  profiles of the vortex solutions of Eqs. \eqref{eq-psi-sol3} and \eqref{eq-psi-sol4} are shown in Fig. \ref{fig:analytic2}. 
 The vortex size can be estimated as $a\sim p^{-1}\sim H$.
\begin{figure*}! 
\begin{center}
\includegraphics[scale=0.4]{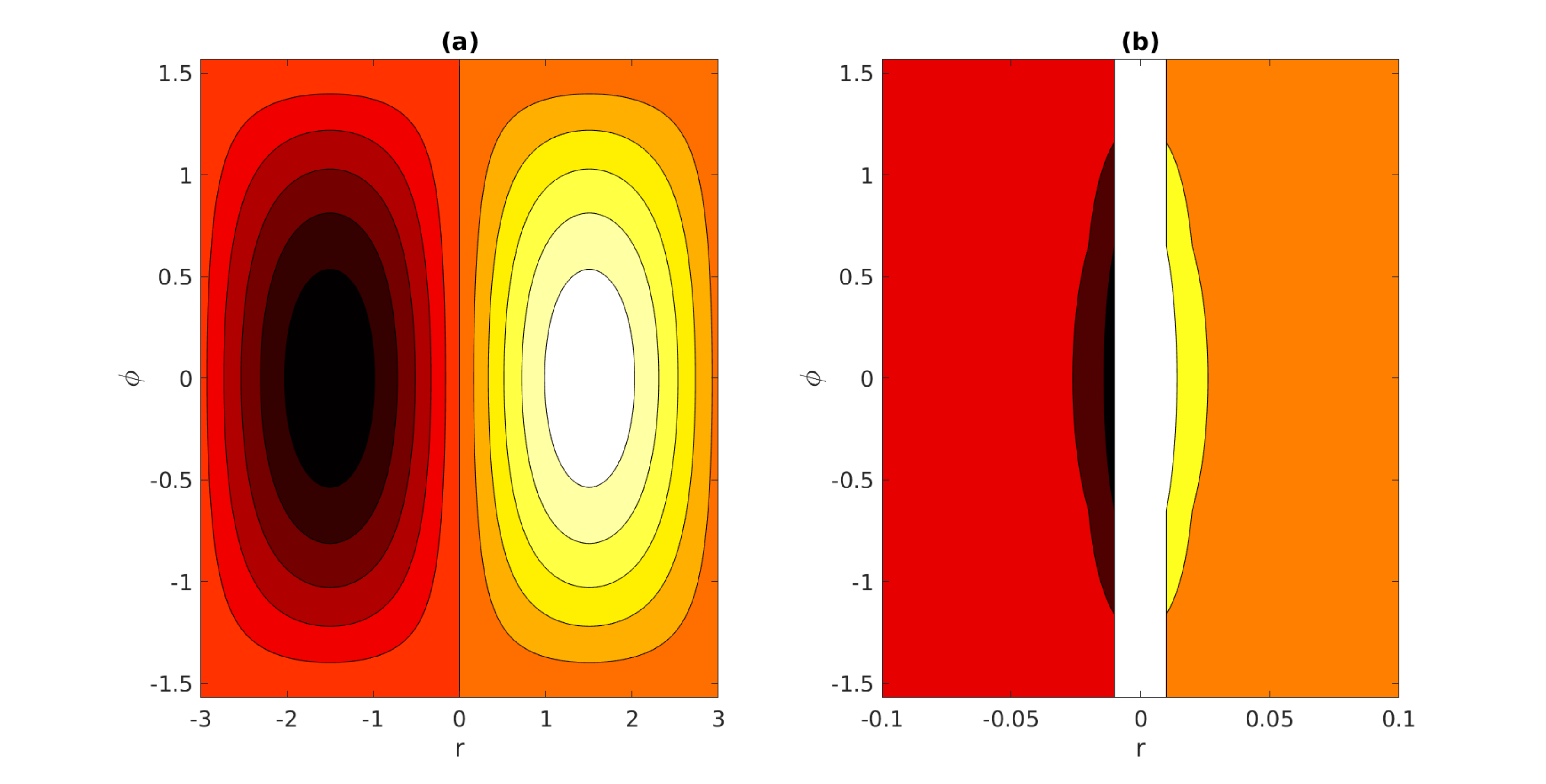}
\caption{Contour plots of  the stationary propagating vortical solutions [(a) Eq. \eqref{eq-psi-sol3} and (b) Eq. \eqref{eq-psi-sol4}] are shown in the $r\phi$ plane. The parameter values are $U/H\omega_g=-23$ for Eq. \eqref{eq-psi-sol3} and $U/H\omega_g=14.1$ for Eq. \eqref{eq-psi-sol4}. The other parameters are $A=0.2$ and $B=2$.}
\label{fig:analytic2}
\end{center}
\end{figure*}
\subsection{An internal gravity dipole vortical solution with the effects of magnetic viscosity $(\eta_0\neq0)$}  \label{sec-vortex-sol3}  
On the basis of the multiple-scale analysis, we discuss the   effects of the magnetic inhibition  (The parameter $\eta_0$ may be called as the magnetic viscosity.  The reason is that in the upper atmosphere with high latitudes, the geostrophic characters of fluid motions are almost lost, and they behave as in a viscous medium. Such viscosity, albeit similar to the magnetic viscosity, however,  appears due to  the inductive inhibition associated with the gyromotion of electrons and ions,  electron-ion collisions, as well as  collisions of neutrals with electrons and ions)  on the vortical solution as described in Sec. \ref{sec-vortex-sol2}. To this end, we rewrite Eq. \eqref{eq-psi11} with account of the magnetic inhibition term $\propto\eta_0=\sigma_pB_0^2/\rho_0$ as
\begin{equation}
\frac{1}{4H^2}\frac{\partial \psi}{\partial t}+\omega_g H\frac{\partial }{\partial x}\nabla^2\psi+\frac{\omega_g}{H}\frac{\partial \psi}{\partial x}-\eta_0\left(\frac{\partial^2}{\partial z^2}-\frac{1}{4H^2}\right)\psi=0. \label{eq-psi44}
\end{equation}  
Assuming that the effect of the magnetic inhibition on the wave damping is small, i.e.,  $\eta_0\sim o(\epsilon)$, we introduce a slow time scale $t_1=\eta_0 t$ and represent $\psi$ as 
\begin{equation}
\psi(t)=\psi^{(1)}(t,t_1)+\eta_0\psi^{(2)}(t,t_1)+{\cal O}(\eta_0^2), \label{eq-pert}
\end{equation}
which applies for longer times, i.e., $t\sim {\cal O}(1/\eta_0)$.
Substituting the perturbation expansion \eqref{eq-pert} into Eq. \eqref{eq-psi44}, we obtain the following first and second order equations.
\begin{equation}
\frac{1}{4H^2}\frac{\partial \psi^{(1)}}{\partial t}+\omega_g H\frac{\partial}{\partial x}\nabla^2\psi^{(1)}+\frac{\omega_g}{H}\frac{\partial \psi^{(1)}}{\partial x}=0, \label{eq-1st-order}
\end{equation}
\begin{equation}
\begin{split}
\frac{1}{4H^2}\frac{\partial \psi^{(2)}}{\partial t}+\omega_g H\frac{\partial}{\partial x}\nabla^2\psi^{(2)}&+\frac{\omega_g}{H}\frac{\partial \psi^{(2)}}{\partial x}=-\frac{1}{4H^2}\frac{\partial \psi^{(1)}}{\partial t_1}\\
&-\frac{1}{4H^2} \psi^{(1)} +\frac{\partial^2 \psi^{(1)}}{\partial z^2}. \label{eq-2nd-order}
\end{split}
\end{equation}
Equation \eqref{eq-1st-order} for $\psi^{(1)}(t,t_1)$ describes the solutions \eqref{eq-psi-sol3} and \eqref{eq-psi-sol4} with the coefficients  $A(t_1)$ and $B(t_1)$ respectively. The secular terms on the right-hand side of Eq. \eqref{eq-2nd-order} (except the last term on the right-hand side, i.e., the source term) can be prevented by imposing the following solvability condition.
\begin{equation}
\frac{\partial \psi^{(1)}}{\partial t_1}+\psi^{(1)}=0. \label{eq-solv-cond}
\end{equation} 
A solution of Eq. \eqref{eq-solv-cond} modulates the coefficients $A(t_1)$ and $B(t_1)$ as well as the profiles  \eqref{eq-psi-sol3} and \eqref{eq-psi-sol4} as
\begin{equation}
\begin{split}
&\psi=A_0\exp(-\eta_0 t)J_1(kr)\cos\phi,\\
&\chi=B_0\exp(-\eta_0 t)K_1(pr)\cos\phi, \label{eq-psi-sol5}
\end{split}
\end{equation}
where $A_0$ and $B_0$, evaluated at $t=0$, are constants. Clearly, the amplitude of the solitary vortex decays as time goes on by the effects of the magnetic viscosity.
\section{Evolution of internal gravity solitary vortex: Numerical approach} \label{sec-numer}
Before we numerically study the system of Eqs. \eqref{eq-psi} and \eqref{eq-chi} for the evolution of dipolar vortices of internal GWs, it is pertinent  to make the system dimensionless. So, redefining the variables, namely
\begin{equation}
\begin{split}
&(x,z)\rightarrow (x,z)/H,~t\rightarrow t\omega_g,~\psi\rightarrow \psi/\psi_0,~\chi\rightarrow\chi/\chi_0,\\
&\alpha=\psi_0/\omega_gH^2,~\beta=H\chi_0/\omega_g\psi_0,~\eta=\sigma_pB_0^2/\rho_0\omega_g, \label{normlization}
\end{split}
\end{equation} 
we obtain from Eqs. \eqref{eq-psi} and \eqref{eq-chi}  the following reduced evolution equations.
 \begin{equation}
 \begin{split}
  \frac{\partial}{\partial t}\left(\nabla^2 \psi -\frac{1}{4}\psi \right) &+\alpha\left( \frac{\partial \psi}{\partial x}\frac{\partial\nabla^2\psi}{\partial z}-\frac{\partial \psi}{\partial z}\frac{\partial\nabla^2\psi}{\partial x}\right) \\
  &=-\beta\frac{\partial \chi}{\partial x}-\eta\left(\frac{\partial^2 \psi}{\partial z^2} -\frac{1}{4}\psi \right), \label{eq-psi-nond} 
  \end{split}
 \end{equation}
  \begin{equation}
 \frac{\partial \chi}{\partial t} +\alpha\left( \frac{\partial \psi}{\partial x}\frac{\partial\chi}{\partial z}-\frac{\partial \psi}{\partial z}\frac{\partial\chi}{\partial x}\right)=\frac{1}{\beta} \frac{\partial \psi}{\partial x}. \label{eq-chi-nond}
 \end{equation} 
\par
Here,  we note that the dimensionless  parameters $\alpha$, $\beta$ and $\eta$   are, respectively,  associated with the Jacobian nonlinearities for the stream function and density variation; the length $(H)$ and the time $(\omega_g^{-1})$ scales; and the Pedersen conductivity.  In order to investigate the global behaviors of solitary vortices we numerically solve  Eqs. \eqref{eq-psi-nond} and \eqref{eq-chi-nond}  by the standard 4-th order Runge-Kutta scheme with the time step $dt=10^{-4}$, the mesh size $dx=dz\sim0.1$ and the initial condition in the form of a two-dimensional sinusoidal wave. The spatial derivatives were approximated by the centered second-order difference formulas.  Furthermore, in order to study the qualitative behaviors of solitary vortices we have considered the parameter values that are relevant, e.g., in the ionospheric $F$-layer. Also,  $\psi_0\sim\omega_g H^2$,  $\chi_0\sim\omega_g^2/ H$, so that $\alpha,~\beta\sim1$, and $\eta \sim 10^{-2}-10^{-1}$ depending on the magnetic field strength.  
 The results are displayed in Figs. \ref{fig:eta1} to \ref{fig:beta2}. It is observed that  the initial  profile evolves and as time goes the nonlinearities (Jacobian) associated with the stream function and the density variation, and  the vorticity function intervene to form a dipolar structure. However, the vortical structure tends to disappear  when the Pedersen parameter $\eta$ exceeds a critical value. The decay of the wave amplitude becomes faster the larger is the values of $\eta$. 
\par 
 Figure \ref{fig:eta1} shows the profiles of the solitary vortex at different times with the parameters $\alpha=1.09$, $\beta=1.02$ and $\eta=0.18$.  From Fig. \ref{fig:eta1} we find that when the effect of the Pedersen parameter is small, i.e., $\eta=0.18$, the decay rate of the wave amplitude is relatively slow for which the dipolar vortex structure can be seen  for a longer time. However, as its value  is increased from $\eta=0.18$ to  $\eta=0.215$, Fig. \ref{fig:eta2} shows that the vortex structure tends to disappear with time and it completely disappears after time $t=80$ due to the energy loss by the magnetic viscosity.     Keeping the values of $\eta$ and $\beta$  fixed at $0.215$ and $1.02$, as the value of $\alpha$ is increased from $\alpha=1.09$ to $\alpha=1.59$, the nonlinearity intervenes to compensate the energy loss. As a result, the vortex structure prevails even for a longer time [See Fig. \ref{fig:alpha2}]. The similar phenomena can happen with   an increasing   value  of $\beta$, say from $\beta=1.02$ to $\beta=1.52$ keeping others fixed at $\alpha=1.09$ and $\eta=0.215$ [See Fig. \ref{fig:beta2}]. Physically, when the parameter $\beta$ is increased, from the energy expression [Eq. \eqref{eq-ener2}] it follows that the nonlinearity associated with the density perturbation also increases, which in turn increases the wave energy ${\cal E}$ and contributes to the energy loss for a limited period of time. As a result, the wave damping becomes slower and the dipolar structure prevails for relatively a longer time.  
\begin{figure*}! 
\begin{center}
\includegraphics[scale=0.35]{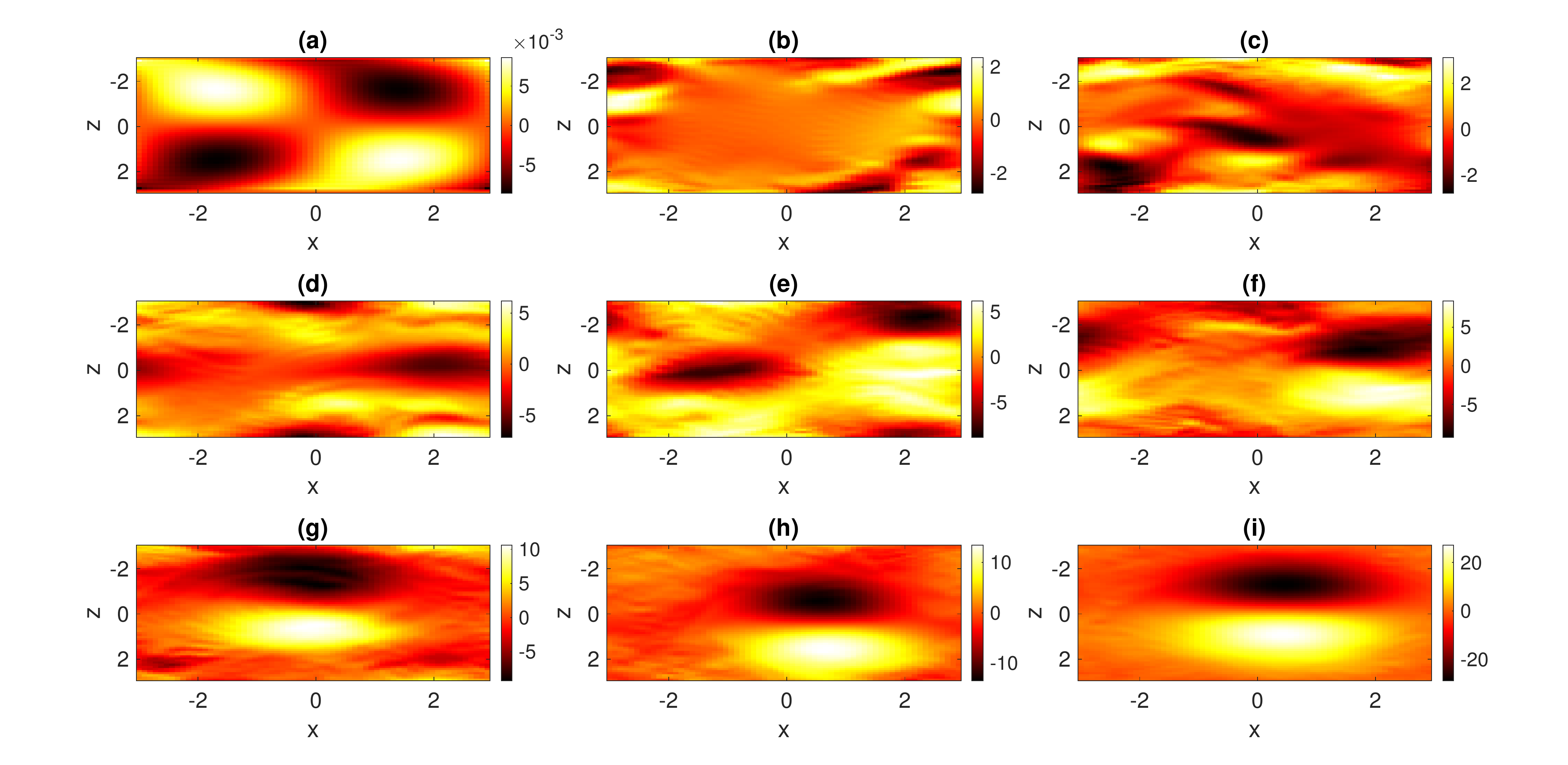}
\caption{ The evolution of solitary dipolar vortex [Numerical simulations of Eqs. \eqref{eq-psi-nond} and \eqref{eq-chi-nond}] is shown at different times from (a) $t=0$  to  (i) $t=80$ with an interval of $t=10$. The parameter values are $\alpha=1.09$, $\beta=1.02$ and $\eta=0.18$.}
\label{fig:eta1}
\end{center}
\end{figure*}
\begin{figure*}! 
\begin{center}
\includegraphics[scale=0.32]{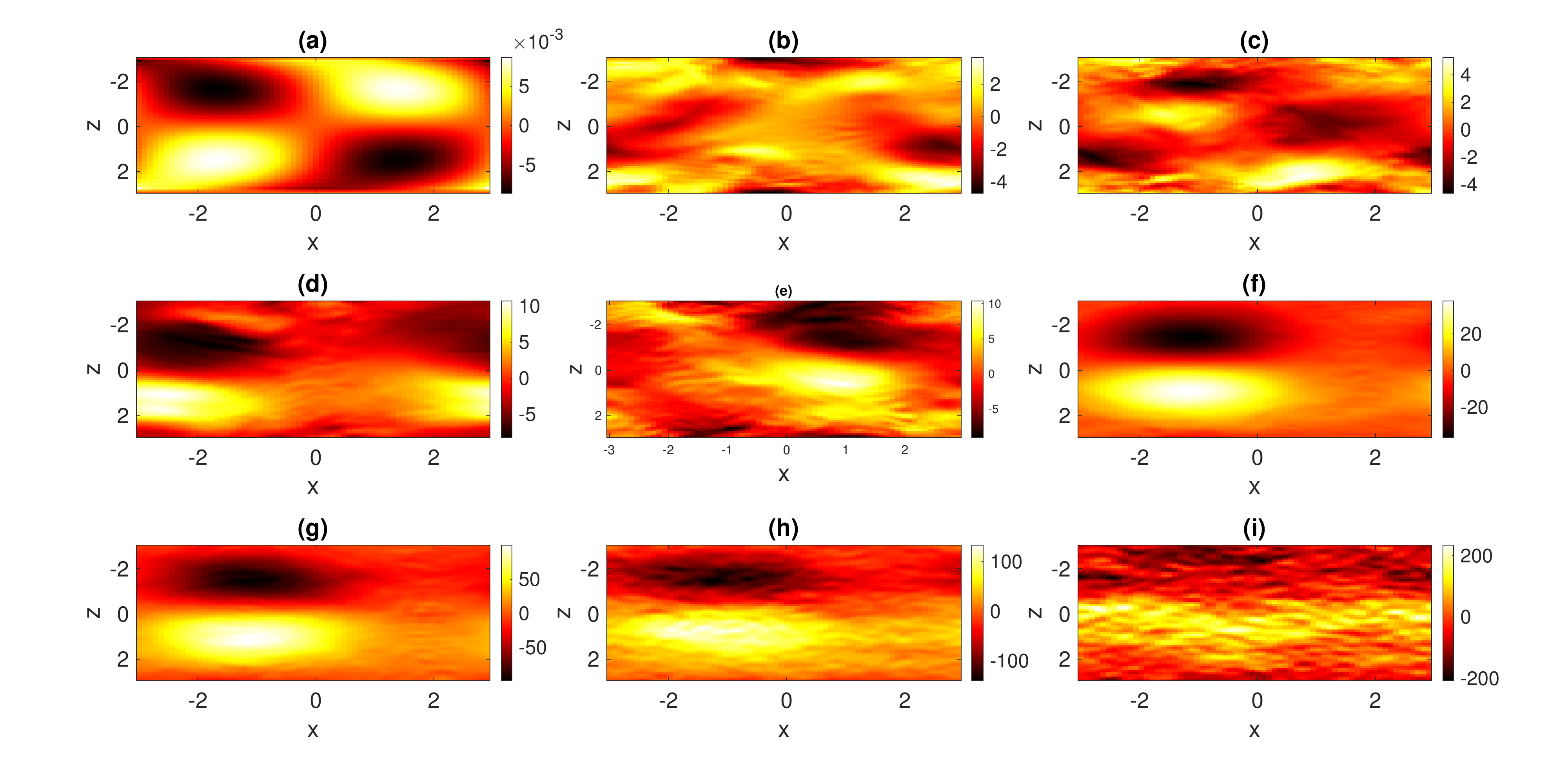}
\caption{The same as in Fig. \ref{fig:eta1} but with a different value of $\eta=0.215$.    The other parameter values are $\alpha=1.09$ and $\beta=1.02$.}
\label{fig:eta2}
\end{center}
\end{figure*}
\begin{figure*}! 
\begin{center}
\includegraphics[scale=0.32]{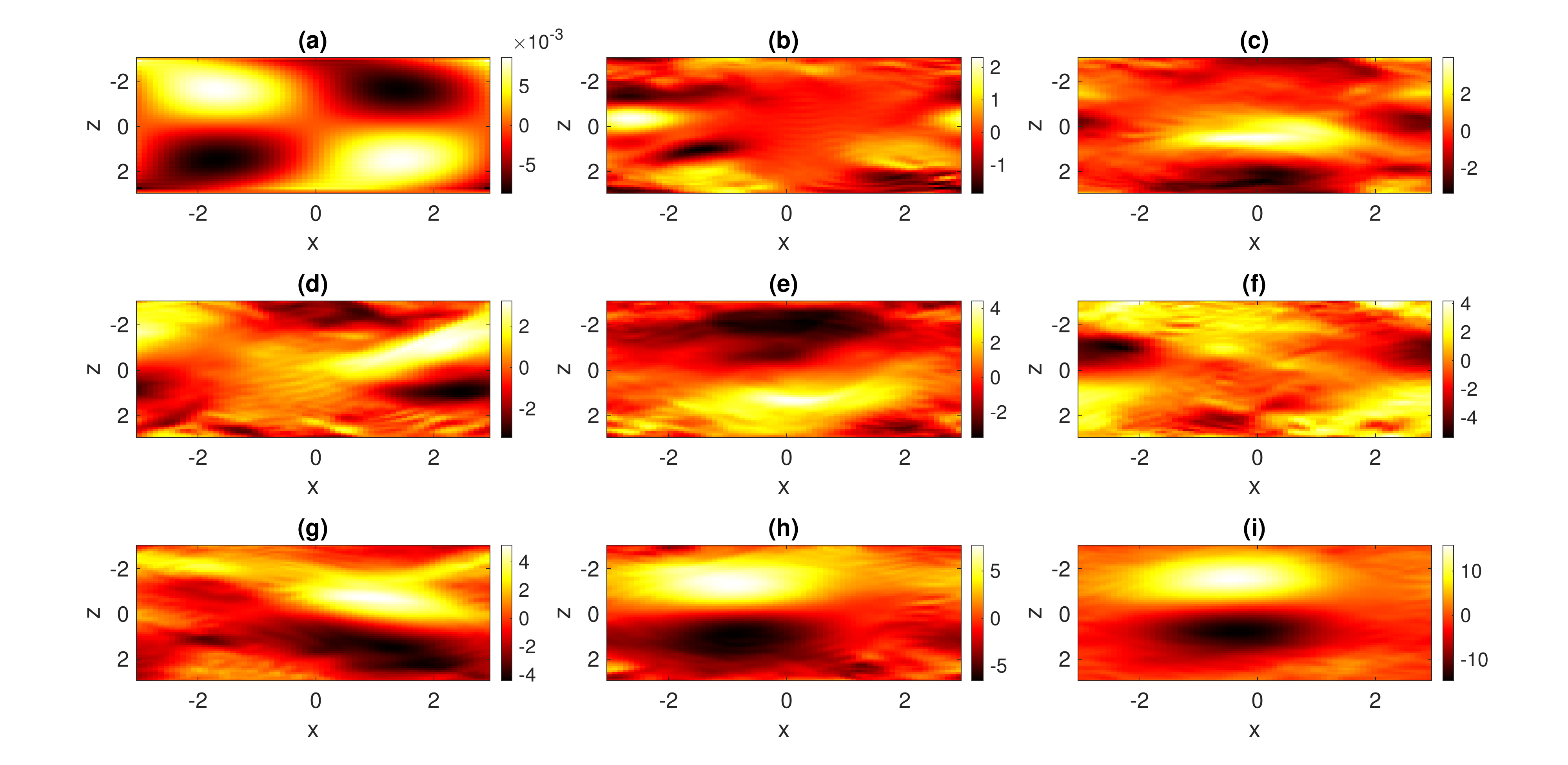}
\caption{ The same as in Fig. \ref{fig:eta2} but with a different value of $\alpha=1.59$.    The other parameter values are $\beta=1.02$ and $\eta=0.215$.}
\label{fig:alpha2}
\end{center}
\end{figure*}
\begin{figure*}! 
\begin{center}
\includegraphics[scale=0.32]{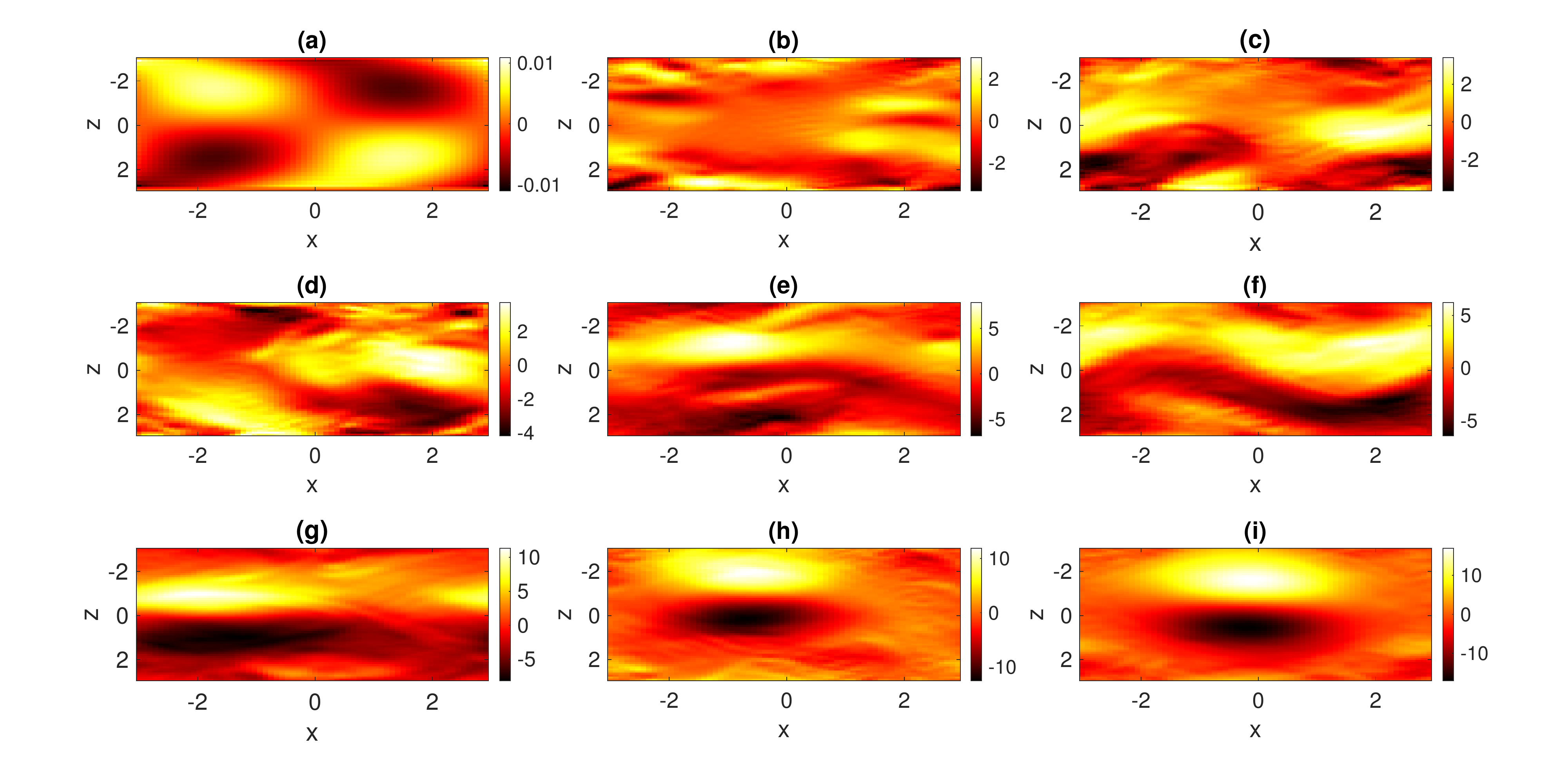}
\caption{ The same as in Fig. \ref{fig:alpha2} but with a different value of $\beta=1.52$.    The other parameter values are $\alpha=1.09$ and $\eta=0.215$.}
\label{fig:beta2}
\end{center}
\end{figure*}
\section{Discussion and Conclusion}\label{sec-conclu} 
 Starting from a set of nonlinear coupled equations \citep{kaladze2008a} for the stream function and the density variations of atmospheric fluids,  we have studied  the nonlinear evolution of    internal  GWs in the Earth's weakly ionized stratified conductive ionosphere.  The  model equations are based on some physical  assumptions.    According to observations, the internal GWs exist in a large range of heights extending from the mesosphere to the ionosphere ($50-500$ km). At such heights, the Earth's ionosphere (overlapping into the mesosphere and thermosphere) is conductive due to the presence of charged particles and neutral atoms. In the model, we have considered the current arising in the gas but neglected the vortex parts of the self-generated electromagnetic fields using the so-called non-inductive approximation. Thus, only the dynamo electric field comes into the picture.  Specifically, we have focused on the nonlinear dynamics of internal GWs  at high latitudes in the northern hemisphere  assuming that the geomagnetic field   is vertical and directed downward. The ionospheric gas is highly stratified  and accordingly, we have considered an isothermal  incompressible atmosphere for the adiabatic  propagation of internal GWs. 
 \par 
    In the small amplitude limit (linear approximation), we have obtained the dispersion relation for internal GWs propagating along the $x$-axis with a frequency lower than than the Brunt-V{\"a}is{\"a}l{\"a} frequency $\omega_g$.   It is seen that  the eigen frequency varies in the regime  $10^{-4}$ s$^{-1}<\omega_0<1.7\times10^{-2}$ s$^{-1}$ and the typical wavelength  is $\lambda\sim H\sim10$ km. Also, the phase velocity and the group velocity are of the same order, i.e., $v_p\sim v_g\sim V_\text{max}\sim6\times10^2$ m/s. Such an estimate agrees with some existing observations made by {\v S}indel{\'a}{\~{r}ov{\'a} \textit{et al.} \citep{sindelarova2009}. 
Furthermore, it is shown that the acting electromagnetic force is defined by the Pedersen conductivity $(\sigma_p)$ leading to the so-called magnetic inhibition (damping) of internal GWs with the decrement rate proportional to the Padersen parameter $\eta\equiv \sigma_pB_0^2/\rho_0\omega_g$. This damping rate increases with the magnetic field strength and for parameter values relevant for the ionospheric $F$-layer.   Its value can be estimated as $|\gamma_0|\sim \eta_0\sim10^{-4}$ s$^{-1}$.       
Due to the small damping rate compared to the eigen frequency $\omega_0\lesssim \omega_g\sim10^{-2}$  s$^{-1}$, the linear internal GWs can grow to a nonlinear stage and form vortical structures whose evolution    is described by a coupled set of nonlinear equations \eqref{eq-psi-nond} and \eqref{eq-chi-nond}. 
 The coupled set of equations is solved  analytically to obtain various vortical solutions with different space localization. Furthermore, the equations are also solved numerically to ascertain the occurrence of internal gravity dipolar vortices.   The internal GWs and related nonlinear structures are widely observed in the upper and lower layers of the atmosphere as well as in the lower ionosphere, (see for details, e.g.,  \citep{fritts2003,swenson1995,picard1998,lu2009,mitchell1998,sindelarova2009}. Also, such   solitary dipolar vortices were discovered in Freja satellite observations of space plasmas \citep{louarn1994,wu1996}. In this context, the differences of the numerical approach implemented here with  some other models, e.g., \citep{gavrilov2020,hickey2015} can be noted. Gavrilov \textit{et al.}  \citep{gavrilov2020}  used     the high-resolution model ``AtmoSym" for three-dimensional simulation of  acoustic GWs with
  background vertical profiles of temperature, molecular weight, density, molecular viscosity and heat conduction corresponding to various solar activity (SA) levels. Their  numerical modeling revealed large  amplitude  temperature perturbations at heights above $150$ km with high SA.       On the other hand, a full wave model was proposed for a binary gas by Hickey \textit{et al.}  \citep{hickey2015} to study the effects of  thermal conductivity and viscosity on the propagation of GWs in the thermosphere.  
\par 
The following new results are  obtained. 
\begin{itemize}
\item For a particular choice of the  mass density $\chi$ as a function of the stream function $\psi$, i.e.,   $\chi=-(\omega_g^2/U)\psi$,  we have obtained a new kind of stationary propagating $(\eta_0=0)$ standard dipole vortical solution \eqref{eq-psi-sol}. The vortical structures can propagate in both the positive (to the East) and negative (to the West) directions of  the $x$-axis with the supersonic velocity $U$, satisfying  $U^2>V^2_\text{max}$ and thereby preventing the generation of linear waves by the moving structure (since the phase velocity is limited by $V_\text{ph}\leq V_\text{max}$). Typical size of the vortex is $a\sim2H\sim15$ km. Furthermore, the vortex structures are exponentially localized in space and regular at the center. 
\item  In   an another case with $\chi=-\omega_g H\nabla^2\psi$, we have obtained another class of  stationary propagating $(\eta_0=0)$ vortical solutions [Eqs. \eqref{eq-psi-sol3} and \eqref{eq-psi-sol4}]. The positive velocity $U$ of the structure [Eq. \eqref{eq-psi-sol3}] is found to be  larger than $V_\text{max}$ and it may vary in the narrow regime $V_\text{max}<U<2V_\text{max}$. Thus, only the structures with negative velocities can   mainly appear. It is found that  the vortex solution is regular at the center $r=0$  and it vanishes  at infinity $(\propto 1/\sqrt{kr})$.  The vortex size can be estimated as $a\sim k^{-1}\sim H\sim10$ km. On the other hand, the vortical structure corresponding to the the solution \eqref{eq-psi-sol4} can propagate with the positive velocity $U>4H\omega_g\sim2V\text{max}$, while the negative velocities (motions to the West) are inhibited.    This dipolar vortex solution has the singularity at the center $r=0$, but is exponentially localized in space  $[\propto\exp(-pr)]$.  The vortex size can be estimated as $a\sim p^{-1}\sim H\sim10$ km.
\item Using the multiple-scale method   approximate vortical solutions \eqref{eq-psi-sol5} of Eqs. \eqref{eq-psi} and \eqref{eq-chi} are also obtained with $\chi=-\omega_g H\nabla^2\psi$ by the influence of weak magnetic inhibition (viscosity).  It is shown that  the amplitude of the solitary vortex profile  decays exponentially with time, i.e., the internal GWs undergo weak damping due to  a small effect of the Pedersen parameter. Such vortical structures exist during the time $t\sim1/\eta_0$ which constitutes $t\sim 30-3\times10^6$ h for the $E$-layer and $30-300$ h for the $F$-layer.
\item Numerical simulations of dimensionless Eqs. \eqref{eq-psi-nond} and \eqref{eq-chi-nond} reveal that the solitary dipolar vortex profile can appear and propagate for a longer time until the dissipation due to the Pedersen conductivity $(\sim\eta\equiv \eta_0/\omega_g)$ is negligible. When the effect of $\eta$ is relatively strong, the vortex structure tends to disappear  as time goes on and it completely disappears after a certain time  due to the energy loss by the magnetic viscosity.  However, the energy loss can reduced or the vortex structure can prevail for relatively a longer time if the nonlinear effects associated with either  the stream function or the density variation or both become significantly higher than that due to the dissipation.  
\end{itemize}
\par 
To conclude, the theoretical results should be useful for understanding the characteristics of low frequency internal gravity waves and the formation of solitary dipolar vortex in the Earth's ionosphere. Such vortical structures  carry trapped particles and contribute to essentially   transport of plasma particles in atmospheric conducting fluids. Thus, they can play crucial roles for understanding  coupling  of the ionosphere with lower-lying regions.   
\section*{Acknowledgement} {One of us (APM) thanks Professor Lennart Stenflo of Link{\"o}ping University, Sweden for his constant encouragement and valuable advice to work in the particular field. APM also acknowledges   support from Science and Engineering Research Board (SERB) for a project with sanction order no. CRG/2018/004475 dated 26 march 2019.   D. Chatterjee acknowledges support from  SERB   for a national postdoctoral fellowship (NPDF)  with sanction order no. PDF/2020/002209 dated 31 Dec 2020. The authors thank Rupak Mukherjee of Princeton University, USA for his useful suggestions to perform the numerical simulation.  }
\bibliographystyle{apsrev4-1}
\bibliography{Reference}
%
%
\end{document}